\documentclass{PoS}
\pdfoutput=1
\usepackage{rotating,booktabs,amsmath,amssymb, multicol}
\usepackage[numbers,sort&compress]{natbib}
\usepackage{natbibspacing}
\newcommand{\comment}[1]{}

\providecommand{\abs}[1]{\lvert#1\rvert}    

\title{Calculating $B$-meson decay constants using domain-wall light quarks and nonperturbatively tuned relativistic $b$-quarks}

\ShortTitle{Calculating $B$-meson decay constants}

\author{\speaker{Oliver Witzel}\\%
        Center for Computational Science, Boston University,\newline 3 Cummington Mall, Boston, MA 02215, USA\\
        E-mail: \email{owitzel@bu.edu}}

\abstract{We calculate $B$-physics quantities using the RBC/UKQCD 2+1 flavor domain-wall plus Iwasaki lattices and the relativistic heavy quark action developed by Christ, Li and Lin. After tuning these parameters nonperturbatively, we present our preliminary results for the calculation of the decay constants $f_B$ and $f_{B_s}$ analyzing data at two lattice spacings of $a \approx 0.11$ fm and $a \approx 0.08$ fm. 
}

\FullConference{The 30th International Symposium on Lattice Field Theory\\
   June 24 -- 29,  2012\\
    Cairns, Australia}

\begin{document}

\section{Introduction}
Using lattice QCD to compute the nonperturbative contributions to $B$-physics quantities allows us to strengthen constraints on the CKM unitarity triangle and determine CKM matrix elements. In the game of over-constraining the apex of the CKM unitarity triangle, $B$-physics may also reveal signs of physics beyond the Standard Model. The standard global fit of the CKM unitarity triangle uses $B$-meson decay constants and mixing as an input \cite{Charles:2004jd,Bona:2005vz,Laiho:2009eu}.\footnote{For~the~latest~results~see: \texttt{http://ckmfitter.in2p3.fr/}, \texttt{http://utfit.roma1.infn.it/} and \texttt{www.latticesaverages.org}.} Experiments such as BaBar, Belle, and LHCb measure $B_q-\overline{B}_q$ mixing in terms of mass differences (oscillation frequencies) $\Delta M_q$ to subpercent accuracy \cite{Beringer:2012zz}. In the Standard Model, these are parameterized by \cite{Buras:1990fn}
\begin{align}
\Delta M_q = \frac{G_F^2m^2_W}{6\pi^2} \eta_B S_0 M_{B_q}{f_{B_q}^2B_{B_q}} \abs{V_{tq}^*V_{tb}}^2,
\end{align}
where the index $q$ denotes a $d$- or a $s$-quark and $M_{B_q}$ the mass of the $B_q$-meson. Perturbatively computed inputs are the QCD coefficient $\eta_B$ \cite{Buras:1990fn} and the Inami-Lim function $S_0$ \cite{Inami:1980fz}; contributions to be computed nonperturbatively are the leptonic $B$-meson decay constant $f_{B_q}$ and the bag parameter $B_{B_q}$.  Precise theoretical determinations of $f^2_{B_q} B_{B_q}$ are needed for precise extraction of the product of CKM matrix elements $V^*_{tq}V_{tb}$. It is in particular advantageous to compute the ratio \cite{Bernard:1998dg}
\begin{align}
\frac{\Delta M_s}{\Delta M_d} = \frac{M_{B_s}}{M_{B_d}}\,{\xi^2} \, \frac{\abs{V_{ts}}^2}{\abs{V_{td}}^2},
\end{align}
because statistical and systematic uncertainties largely cancel and the nonperturbative contribution is contained in the $SU(3)$ breaking ratio
\begin{align}
\xi &= \frac{f_{B_s}\sqrt{B_{B_s}}}{f_{B_d}\sqrt{B_{B_d}}}.
\end{align}
Further, constraints of the CKM triangle from $BR(B\to \tau\nu)$ require precise knowledge of the decay constant $f_B$ \cite{Lunghi:2009ke,Bona:2009cj,Lenz:2010gu}. Currently, only HPQCD and Fermilab/MILC have published results for the decay constants $f_B$ and $f_{B_s}$ as well as $\xi$ using 2+1 dynamical flavor gauge field configurations and ensembles with different lattice spacing. However, both groups use MILC's Asqtad lattices \cite{Gamiz:2009ku,Na:2012kp,Bazavov:2011aa,Bazavov:2012zs} and hence an independent cross-check is desirable.

Simulating $B$-physics on the lattice provides a special challenge because an additional scale given by the large $b$-quark mass needs to be accommodated. Here we report on our project to compute $B$-physics quantities using the nonperturbatively tuned relativistic heavy quark action for $b$-quarks and domain-wall fermions for the light $u,\,d,\,s$-quarks.

\section{Computational setup}
We base our $B$-physics project on gauge-field configurations generated by the RBC/UKQCD collaborations with 2+1 flavors of domain-wall fermions and the Iwasaki gluon action \cite{Allton:2008pn,Aoki:2010dy}. The ensembles and their parameters are listed in Tab.~\ref{tab:lattices}.  Using the domain-wall action \cite{Kaplan:1992bt,Shamir:1993zy} we generate on each configuration six light valence-quark propagators with quark masses $a m_\text{val}^{24}$ = 0.005, 0.010, 0.020, 0.030, 0.0343 and $0.040$ on the coarser $24^3$ ensembles and $a m_\text{val}^{32}$ = 0.004, 0.006, 0.008, 0.025, 0.0272 and 0.030 on the finer $32^3$ ensembles. On the two ``coarse'' $24^3$ ensembles ($a \approx 0.11 $fm; $a^{-1} = 1.729$ GeV) with light sea-quark mass $a m_l=0.005$ and $0.01$ we place one time source per configuration, while on the ``fine'' $32^3$ ensembles ($a \approx 0.086$ fm; $a^{-1} = 2.281$ GeV) with light sea-quark masses $a m_l=0.004,\, 0.006$ and $0.008$ we place two time sources per configuration separated by half the temporal extent of the lattice. The mass of the light quark propagators with $a m^{24}_\text{val} = 0.0343$ and  $a m^{32}_\text{val} = 0.0272$ is extremely close to the value of the physical strange quark mass on the corresponding ensembles: $a m_s^{24} = 0.0348(11)$ and $a m_s^{32} = 0.0273(11)$. 

For the $b$ quarks we use the anisotropic Sheikholeslami-Wohlert (clover) action with the relativistic heavy-quark (RHQ) interpretation~\cite{Christ:2006us,Lin:2006ur}. We require two experimental inputs for tuning the three parameters, $m_0a$, $c_P$, $\zeta$, nonperturbatively. In our notation the RHQ action is defined by
\begin{align}
S = \sum_{n,n^\prime}\bar\Psi_n \left\{\! m_0+ \gamma_0 D_0 -\! \frac{a D_0^2}{2} + \zeta\left[\vec \gamma \cdot \vec D- \frac{a\left(\vec D \right)^2}{2}\right]\!- a \sum_{\mu\nu} \frac{i c_P}{4}\sigma_{\mu\nu}F_{\mu\nu}\!\right\}_{\!\!n,n^\prime}\!\!\!\!\Psi_{n^\prime}.
\end{align}

\begin{table}[t]
\centering
\caption{Lattice simulation parameters used in our $B$-physics program.  The columns list the lattice volume, approximate lattice spacing, light ($m_l$) and strange ($m_h$) sea-quark masses, unitary pion mass, and number of configurations and time sources analyzed.}
\vspace{3mm}
\label{tab:lattices}
\begin{tabular}{cccccrc} \toprule
  &         &         &         &            & &\# time\\
$\left(L/a\right)^3 \times \left(T/a\right)$ \qquad & $\approx a$(fm) & ~~$am_l$ & ~~$am_h$ & \quad $M_\pi$(MeV) \quad & \# configs.&sources\\[0.5mm] \midrule
$24^3 \times 64$ &  0.11 &  0.005 & 0.040 & 329 & 1636~~~~&1\\
$24^3 \times 64$ &  0.11 &  0.010 & 0.040 & 422 & 1419~~~~&1\\ \midrule
$32^3 \times 64$ &  0.086 &  0.004 & 0.030 & 289 & 628~~~~&2\\ 
$32^3 \times 64$ &  0.086 &  0.006 & 0.030 & 345 & 889~~~~&2\\
$32^3 \times 64$ &  0.086 &  0.008 & 0.030 & 394 & 544~~~~&2\\ \bottomrule
\end{tabular}\end{table}

\begin{figure}[htb]
\centering
\includegraphics[scale=0.5]{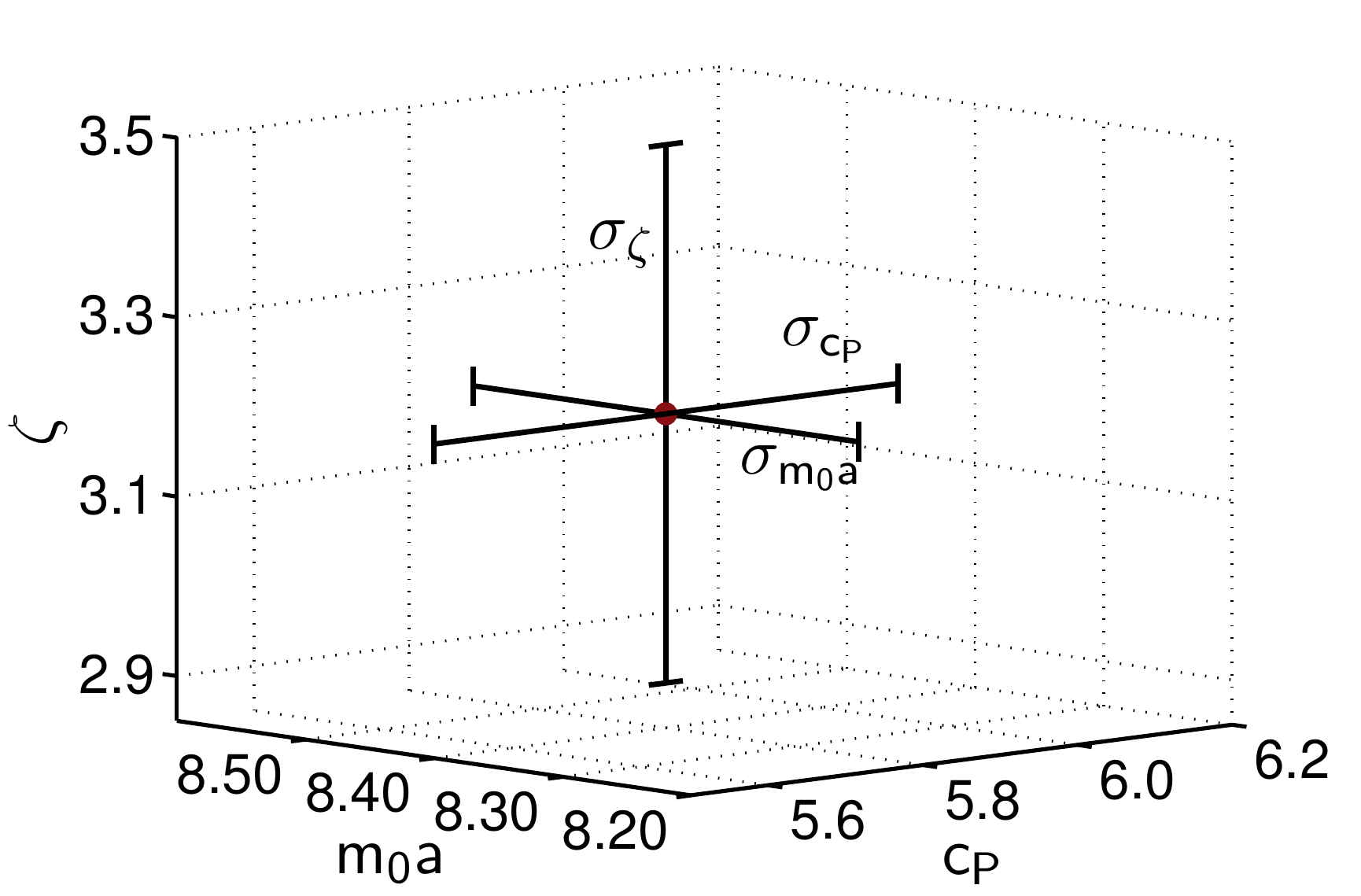}
\caption{The seven RHQ parameter sets are indicated by the six endpoints and the central point of the molecule-like structure in the $m_0a$, $c_P$ and $\zeta$ parameter space.}
\label{Fig:RHQparameters}
\end{figure}

\noindent We tune the parameters by probing seven points of the parameter space (see Fig.~\ref{Fig:RHQparameters}) measuring each time the spin averaged $B_s$-meson mass $\overline M$, the hyperfine splitting $\Delta_M$ and the ratio of the rest mass over the kinetic mass $M_1/M_2$. Assuming linear dependence of these measured quantities on the parameters for the seven points used, we obtain the tuned parameters by matching $\overline M$ and $\Delta_M$ to experimental results and demanding that $M_1/M_2$ equals one. Typically this requires a few iterations to ensure that the tuned parameters are given by an interpolation in a region where the measured parameter dependence is sufficiently linear. Further details are presented in \cite{Aoki:2012xaa} where we also predict bottomonium masses and mass splittings as test of our tuned parameters.

\begin{table}[t]
\centering
\caption{Tuned parameters $m_0a$, $c_P$, $\zeta$ of the RHQ action for simulating $b$-quarks on the ensembles listed in Table I. For details see \cite{Aoki:2012xaa}.}
\label{tab:RHQparameters}
\vspace{2mm}
\begin{tabular}{ccccc}\toprule
$\approx a(fm)$& $m_{sea}^l$ & $m_0a$ & $c_P$ & $\zeta$ \\ \midrule
0.11& 0.005 & 8.43(7) &5.7(2) &3.11(9)\\
0.11& 0.010 & 8.47(9) &5.8(2) &3.1(2)\\ \midrule
0.086& 0.004 & 4.07(6) & 3.7(1) &1.86(8)\\
0.086& 0.006 & 3.97(5) & 3.5(1) &1.94(6)\\
0.086& 0.008 & 3.95(6)& 3.6(1) &1.99(8)\\\bottomrule
\end{tabular}
\end{table}

The computation of the $B$-meson decay constants presented here is performed similarly to the tuning, {\it i.e.}~we compute decay constants at the same seven points in parameter space (see Fig.~\ref{Fig:RHQparameters}) and then interpolate these measurements to the value of the tuned RHQ parameters given in Tab.~\ref{tab:RHQparameters}. We measure the decay constant $f_B$ by computing on the lattice the decay amplitude $\Phi_B$ which is proportional to the vacuum-to-meson matrix element of the heavy-light axial  vector current ${\cal A}_\mu = \bar{b} \gamma_5 \gamma_\mu q$
\begin{align}
\langle 0| {\cal A}_\mu | B(p)\rangle / \sqrt{M_B} = i p^\mu \Phi_B^{(0)} / M_B
\end{align}
with $p_\mu$ the four-momentum of the $B$-meson. To reduce discretization errors we use an $O(a)$-improved axial vector current on the lattice
\begin{align}
\Phi_B^\text{imp} = \Phi_B^{(0)} + c_1 \Phi_B^{(1)},
\end{align}
where the coefficient $c_1$ is computed at 1-loop with mean-field improved lattice perturbation theory \cite{Lehner:TalkLat2012}. 
Multiplying the decay amplitude by the renormalization factor $Z_\Phi$, the lattice spacing $a$ and the mass of the $B_q$-meson $M_{B_q}$ we obtain the decay constant
\begin{align}
f_{B_q} = Z_\Phi \Phi_B^\text{imp} a^{-3/2} / \sqrt{M_{B_q}}.
\end{align}
In our final setup we intend to compute the renormalization factor $Z_\Phi$ following the mostly nonperturbative method outlined in \cite{ElKhadra:2001rv}, {\it i.e.}~$Z_\Phi$ is split into a nonperturbative part containing the flavor-conserving factors $Z_V^{ll}$ and $Z_V^{bb}$ and a perturbatively computed factor $\varrho_{bl}$ which is expected to be close to one and to have a more convergent series expansion in $\alpha_s$:
\begin{align}
Z_\Phi = \varrho_{bl} \sqrt{Z_V^{bb} Z_V^{ll}}.
\end{align} 
Here we use the nonperturbatively obtained $Z_V^{ll}$ \cite{Aoki:2010dy}, while we compute $Z_V^{bb}$ and $\varrho_{bl}$ at 1-loop in mean-field improved lattice perturbation theory \cite{Lehner:TalkLat2012}. Work is in progress to obtain $Z_V^{bb}$ nonperturbatively on all ensembles \cite{Kawanai:2012id}.

\section{Preliminary results}
\begin{figure}[t]
\centering
\includegraphics[scale=0.5]{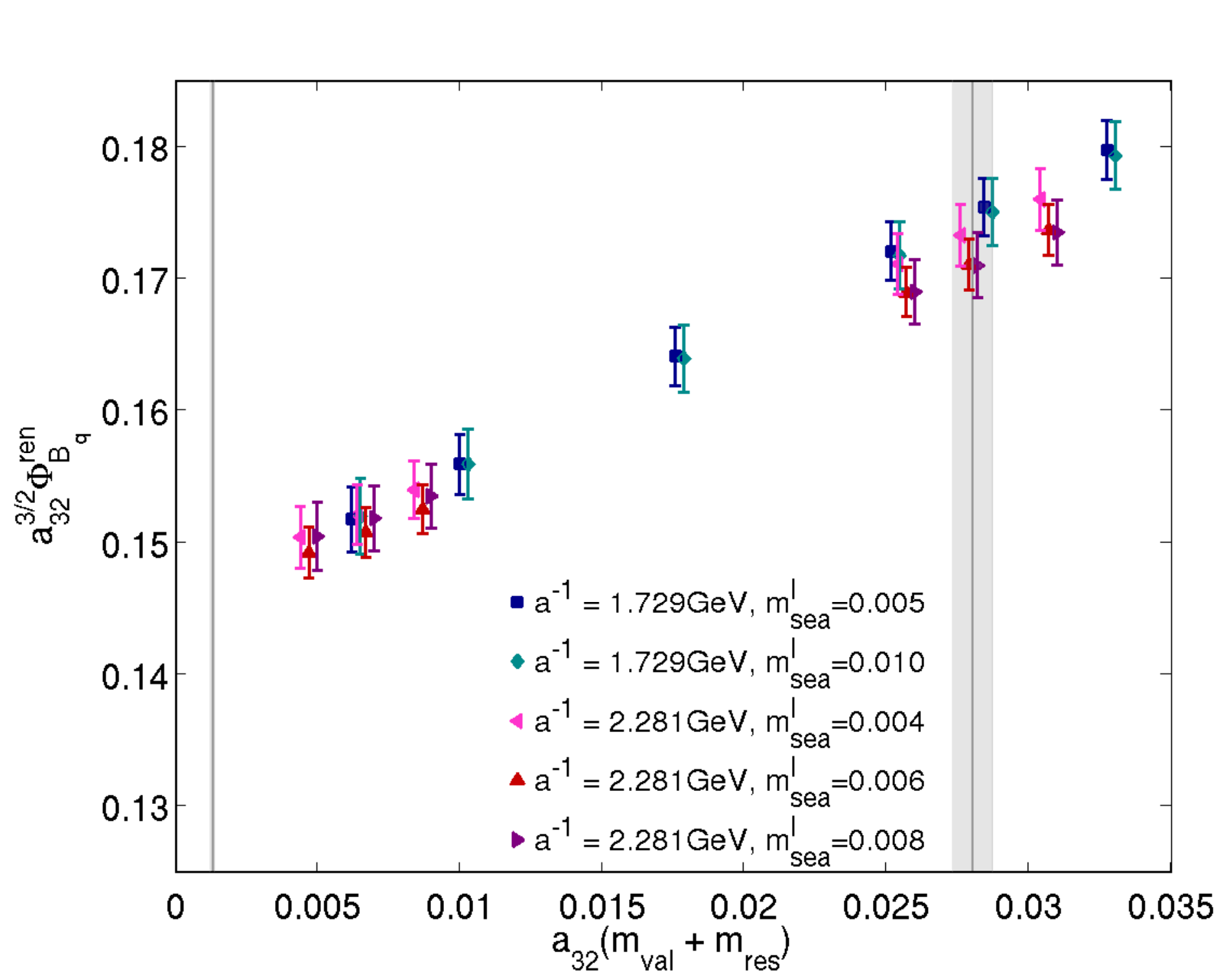}
\caption[width=0.96\textwidth]{Renormalized and $O(a)$-improved decay amplitude $\Phi_{B_q}$ computed on the full data set and shown in $a_{32}$ lattice units. For better visibility data points for $m_\text{sea}^l=0.01$ and $0.008$ are plotted with a small, positive, horizontal offset; data points for $m_\text{sea}^l=0.004$ have a small, negative, horizontal offset. The vertical gray lines with error-band indicate the physical values of the $d$- and $s$-quark masses \cite{Blum:2010ym,Allton:2008pn,Aoki:2010dy}.} 
\label{Fig:PhiB}
\end{figure}

\begin{figure}[t]
\centering
\includegraphics[scale=0.5]{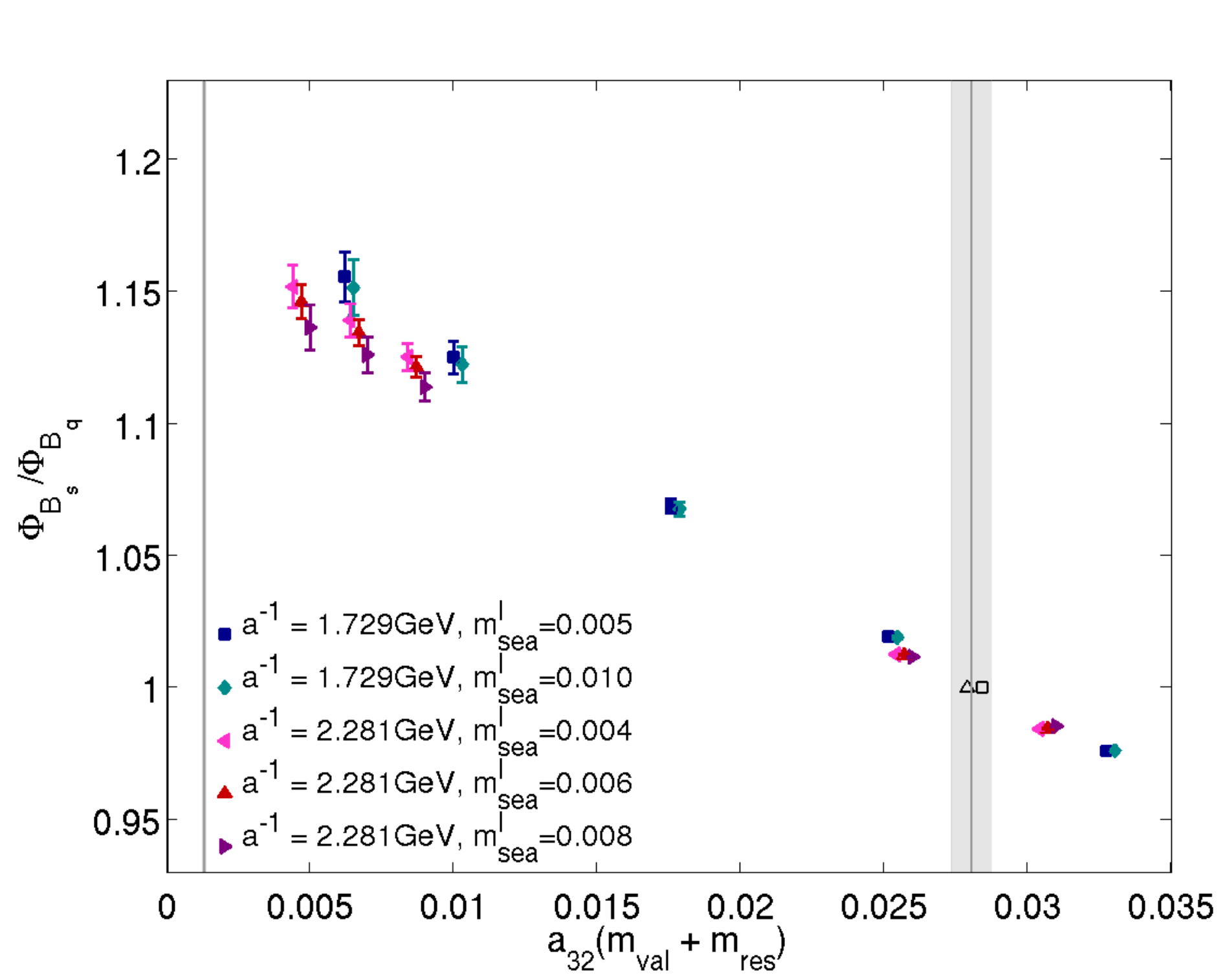}
\caption[width=0.96\textwidth]{Ratio of decay amplitudes $\Phi_{B_s}/\Phi_{B_q}$ computed on the full data set and shown in $a_{32}$ lattice units. For better visibility data points for $m_\text{sea}^l=0.01$ and $0.008$ are plotted with a small, positive, horizontal offset; data points for $m_\text{sea}^l=0.004$ have a small, negative, horizontal offset. The vertical gray lines with error-band indicate the physical values of the $d$- and $s$-quark masses \cite{Blum:2010ym,Allton:2008pn,Aoki:2010dy}. The open black triangle (square) marks the value of the valence strange-quark mass in the simulation at which the plotted ratio is one by construction.}
\label{Fig:PhiB_PhiBs}
\end{figure}

In Fig.~\ref{Fig:PhiB} we show our preliminary results for the renormalized, $O(a)$-improved decay amplitude $\Phi_B^\text{ren} = Z_\Phi \Phi_B^\text{imp}$. While a chiral extrapolation to the physical $d$-quark mass  is needed to obtain $\Phi_{B_d}$, we may take advantage of the simulation data close to the physical strange quark mass to read off a rough estimate of $\Phi_{B_s}$. This indicates our result is in agreement with the existing literature and we expect statistical uncertainties of a few MeV. Figure \ref{Fig:PhiB_PhiBs} shows our results for the SU(3) breaking ratio of $\Phi_{B_s}/\Phi_{B_q}$. The data exhibit a weak dependence on the light sea-quark mass as well as a mild dependence on the lattice spacing. 

\section{Outlook}
Currently work is in progress on the nonperturbative computation of $Z_V^{bb}$ as well as checks on the perturbative computation of the correction factor $\varrho_{bl}$ and the $O(a)$-improvement coefficient $c_1$. We are also working on combined, correlated fits using expressions based on SU(3) and SU(2) heavy-light meson chiral perturbation theory \cite{Goity:1992tp,Arndt:2004bg,Aubin:2005aq,Albertus:2010nm}. With the individual parts being in good shape we hope to publish results soon. We will move on to the computation of $B_q-\overline{B}_q$ mixing matrix elements and their ratio $\xi$.

\section*{Acknowledgments}
We thank our colleagues of the RBC and UKQCD collaborations for useful help and discussions. Numerical computations for this work utilized USQCD resources at Fermilab, in part funded by the Office of Science of the U.S.~Department of Energy, as well as computers at Brookhaven National Laboratroy and Columbia University. O.W. acknowledges  support at Boston University by the U.S. DOE grant DE-FC02-06ER41440.

\clearpage
\bibliography{B_meson}
\bibliographystyle{apsrev4-1}

\end{document}